\documentclass[preprint]{aastex63}

\newcommand{\alfven}{Alfv\'{e}n}

\usepackage{amsmath}
\usepackage{color}
\usepackage{multirow}
\usepackage{tabularx}
\usepackage{diagbox}
\usepackage{tikz}
\usepackage{nicematrix}

\usepackage{natbib} 
\usepackage{bm}
\usepackage{threeparttable}

\begin{document}

\title{Unveiling Spatiotemporal Properties of the Quasi-periodic Pulsations in the Balmer Continuum at 3600 \AA\ in an X-class Solar White-light Flare}

\author[0000-0003-0057-6766]{De-Chao Song}
\affiliation{Key Laboratory of Dark Matter and Space Astronomy, Purple Mountain Observatory, CAS, Nanjing 210023, People's Republic of China}

\author[0000-0002-1196-4046]{Marie Dominique}
\affiliation{Solar-Terrestrial Centre of Excellence -- SIDC, Royal Observatory of Belgium, Brussels, Belgium}

\author[0000-0001-6995-3684]{Ivan Zimovets}
\affiliation{Space Research Institute (IKI), Russian Academy of Sciences, Profsoyuznaya St. 84/32, Moscow, 117997, Russia}

\author[0000-0001-7540-9335]{Qiao Li}
\affiliation{Key Laboratory of Dark Matter and Space Astronomy, Purple Mountain Observatory, CAS, Nanjing 210023, People's Republic of China}
\affiliation{School of Astronomy and Space Science, University of Science and Technology of China, Hefei 230026, People's Republic of China}

\author[0000-0002-8258-4892]{Ying Li}
\affiliation{Key Laboratory of Dark Matter and Space Astronomy, Purple Mountain Observatory, CAS, Nanjing 210023, People's Republic of China}
\affiliation{School of Astronomy and Space Science, University of Science and Technology of China, Hefei 230026, People's Republic of China}

\author[0000-0002-1713-2160]{Fu Yu}
\affiliation{Key Laboratory of Dark Matter and Space Astronomy, Purple Mountain Observatory, CAS, Nanjing 210023, People's Republic of China}

\author[0000-0002-4241-9921]{Yang Su}
\affiliation{Key Laboratory of Dark Matter and Space Astronomy, Purple Mountain Observatory, CAS, Nanjing 210023, People's Republic of China}
\affiliation{School of Astronomy and Space Science, University of Science and Technology of China, Hefei 230026, People's Republic of China}

\author[0000-0002-3410-6706]{B.A. Nizamov}
\affiliation{Space Research Institute (IKI), Russian Academy of Sciences, Profsoyuznaya St. 84/32, Moscow, 117997, Russia}
\affiliation{Sternberg Astronomical Institute, Lomonosov Moscow State University, 119234, Universitetskiy Prospekt, 13, Moscow, Russia}

\author[0000-0003-3699-4986]{Ya Wang}
\affiliation{Key Laboratory of Dark Matter and Space Astronomy, Purple Mountain Observatory, CAS, Nanjing 210023, People's Republic of China}

\author[0000-0003-4490-7344]{Andrea Francesco Battaglia}
\affiliation{Istituto ricerche solari Aldo e Cele Daccò (IRSOL), Università della Svizzera italiana, Locarno, Switzerland}

\author[0000-0002-1068-4835]{Jun Tian}
\affiliation{Key Laboratory of Dark Matter and Space Astronomy, Purple Mountain Observatory, CAS, Nanjing 210023, People's Republic of China}
\affiliation{School of Astronomy and Space Science, University of Science and Technology of China, Hefei 230026, People's Republic of China}

\author[0000-0003-4655-6939]{Li Feng}
\affiliation{Key Laboratory of Dark Matter and Space Astronomy, Purple Mountain Observatory, CAS, Nanjing 210023, People's Republic of China}
\affiliation{School of Astronomy and Space Science, University of Science and Technology of China, Hefei 230026, People's Republic of China}

\author[0000-0003-1078-3021]{Hui Li}
\affiliation{Key Laboratory of Dark Matter and Space Astronomy, Purple Mountain Observatory, CAS, Nanjing 210023, People's Republic of China}
\affiliation{School of Astronomy and Space Science, University of Science and Technology of China, Hefei 230026, People's Republic of China}

\author[0000-0001-9979-4178]{W. Q. Gan}
\affiliation{Key Laboratory of Dark Matter and Space Astronomy, Purple Mountain Observatory, CAS, Nanjing 210023, People's Republic of China}
\affiliation{University of Chinese Academy of Sciences, Nanjing 211135, People's Republic of China}

\correspondingauthor{De-Chao Song}
\email{dcsong@pmo.ac.cn}

\begin{abstract}
Quasi-periodic pulsations (QPPs) in the Balmer continuum of solar white-light flares (WLFs) are rarely reported, and accurately pinpointing the spatial source of flaring QPPs remains a significant challenge.
We present spatiotemporal characteristics of QPPs of an X2.8 two-ribbon solar WLF (SOL2023-12-14T17:02), which was well observed by the White-light Solar Telescope (WST) aboard the Advanced Space-based Solar Observatory, with high-cadence imaging (1--2 s) in the Balmer continuum at 3600 \AA. Combined with additional multi-instrument data, we find that the enhancement of the WLF in both Balmer and Paschen continua shows strong spatiotemporal correlation with hard X-ray (HXR) emissions.
Notably, the pulses in the WST Balmer continuum exhibited a near-zero time lag with most HXR pulses, whereas soft X-ray and extreme ultraviolet emissions showed a lag of 2--3 s.
Interestingly, quasi-harmonic QPPs with periods of $\sim$11 s and $\sim$20 s were observed in multiple wavelengths in the rising phase of the white-light continuum.
Furthermore, we employed Fourier transform to spatially locate the QPPs around 11 and 20 s, revealing that they primarily originated from the east flare ribbon, which exhibited the most substantial continuum enhancement.
More interestingly, we find that the west ribbon contributed significantly to the 11-second QPP but had a weaker contribution to the 20-second QPP. 
Moreover, the occurrence of quasi-harmonic QPPs is temporally coincident with the rapid elongation and separation motions of flare ribbons. Possible mechanisms for the quasi-harmonic QPPs have been discussed.
These observations provide valuable insights into QPP modeling for solar and stellar flares.

\end{abstract}


\section{Introduction} 
\label{sec:intro}

Quasi-periodic pulsations (QPPs) are characterized by quasi-periodic intensity variations in electromagnetic radiation over time \citep{McLaughlin2018}. 
They are widely observed in both solar and stellar flares and closely associated with key physical processes, such as magnetic reconnection, magnetohydrodynamic (MHD) waves, and particle acceleration \citep{Zimovets2021,Ruan2019}. QPPs span nearly the entire electromagnetic spectrum, with timescales ranging from sub-seconds to several tens of minutes \citep[e.g.][]{McAteer2005,Zimovets2010,Nakariakov2010,Simoes2015,ZhaoXZ2019,Lu2021,Hayes2020,Guo2023,Inglis2024,LD2025AA}. 

The study of QPPs in solar flares has a history of over fifty years. The research primarily relies on full-disk integrated flux data in X-ray and radio wavelengths \citep[e.g.][]{InglisNakariakov2009,Milligan2017,Inglis2024}, low-cadence spatially resolved imaging \citep[e.g.][]{Yuan2019}, or high-cadence spectral observations within a limited field of view \citep[e.g.][]{Mariska2006,Tian2016}. Recently, several studies on flaring QPPs have also been reported based on high-cadence radio imaging \citep[e.g.][]{Kou2022,Chen2019}.
By integrating various theoretical calculations and numerical simulations, at least 15 mechanisms have been proposed to explain QPPs, as summarized in the review by \citet{Zimovets2021}. These mechanisms have significantly advanced our understanding of the physical processes underlying QPPs in flares. 
However, several open questions remain \citep[see][]{Inglis2023BAAS}. 
For example, where are the sources of QPPs within flare structures \citep{Nakariakov2003,Melnikov2005,Inglis2008,Kupriyanova2013}? 
What are the fundamental differences and connections between QPPs observed in solar and stellar flares \citep{McLaughlin2018,Zimovets2021}? 
Research on QPPs in solar white-light flares (WLFs) can improve our understanding of these issues, as WLFs can offer a closer analogy to stellar flares due to substantial energy releases \citep{Carrington1859}.

QPPs in the white-light band are frequently observed in stellar flares \citep[e.g.,][]{Mathioudakis2006,Tsap2011} but rarely reported in solar flares \citep{McAteer2005}, particularly in the white-light continuum. 
Studies of QPPs in solar WLFs can offer critical references for understanding their stellar counterparts. 
Very recently, \citet{LD2024} identified an 8.6-minute flare QPP in the white-light continuum at 6173 \AA, likely modulated by the slow-mode magnetoacoustic gravity wave leaking from the sunspot penumbra.
A similar QPP period (approximately 8 mins) was also reported in flare loops in the same white-light continuum by \citet{Zhao2021}, probably caused by pulsive magnetic reconnection.
Undoubtedly, high-cadence observations of WLFs can uncover more finer details of energy release and deposition processes in QPPs.  

The recently launched Advanced Space-based Solar Observatory (ASO-S; \citealt{Gan2023}) enables high-cadence imaging (1--2 s) of solar flares in the Balmer continuum at 3600 \AA, providing critical diagnostics of the radiative characteristics of WLFs in the middle-to-lower chromosphere and corona \citep{hao17nc,Jing2024,LY2024a,LY2024b}. 
In this study, we investigate the spatiotemporal properties of QPPs in the 3600 \AA\ continuum of an X2.8 WLF.
For the first time, we reveal quasi-harmonic QPPs in the 3600 \AA\ continuum of a solar WLF and spatially locate their sources.

\section{Instruments and data reduction}
\label{instr_data}

The X2.8 flare was comprehensively observed by multiple telescopes (see details in Table \ref{tab1}), including the Lyman-alpha Solar Telescope (LST; \citealt{Feng2019,LiH2019,Chen2024}), which contains the White-light Solar Telescope (WST), and the Hard X-ray Imager (HXI; \citealt{Su2024}), both onboard the ASO-S, the Helioseismic and Magnetic Imager (HMI; \citealt{HMI2012}) and the Atmospheric Imaging Assembly (AIA; \citealt{AIA2012}) onboard the Solar Dynamics Observatory (SDO; \citealt{SDO2012}), the Spectrometer Telescope for Imaging X-rays (STIX; \citealt{Krucker2020}) aboard the Solar Orbiter (SolO; \citealt{Muller2020}), the Expanded Owens Valley Solar Array (EOVSA; \citealt{Gary2018}), the Large-Yield RAdiometer (LYRA; \citealt{Dominique2013}) onboard the PRoject for On-Board Autonomy 2, the Gamma-ray Burst Monitor (GBM; \citealt{Meegan2009}) on Fermi, the X-Ray Sensor (XRS; \citealt{XRS1996}) on Geostationary Operational Environmental Satellite (GOES).

\begin{table}[htbp]
  \centering
  \caption{Information on the Instruments and Data Used in This Study}
    \begin{tabular}{lccc}
		\hline
		\hline
        Instrument & Waveband & Spatial Resolution & Cadence (Nominal/Used) \\ \hline
        ASO-S/LST/WST & 3600$\pm$20 \AA\  & 3\arcsec & 1--2 s or 2 mins/1--2 s  \\ 
        ASO-S/HXI & 10--300 keV & - & 0.125 s/0.25 s \\ \hline
        SDO/HMI & 6173 \AA\ & 1\arcsec & 45 s/45 s \\ 
        SDO/AIA & 1600 \AA\ & 1.2\arcsec & 24 s/24 s\\ 
        ~ &              131 \AA\ & 1.2\arcsec & 12 s/12 s \\ \hline
        SolO/STIX & 4--150 keV & - & 0.1 s/0.5 s \\ \hline
        Fermi/GBM & 8 keV--40 MeV & - & 0.064 s or 1 s/1 s \\ \hline
        GOES-16/XRS & 0.5--4 \AA\ & - & 1 s/1 s \\ 
        ~                    & 1--8 \AA\ & - & 1 s/1 s \\ \hline
        PROBA2/LYRA & 1--200 \AA\ & - & 0.05 s/0.05 s \\ 
        ~                    & 1--800 \AA\ & - & 0.05 s/0.05 s\\ \hline
        EOVSA & 1--18 GHz & $\geq$6\arcsec & 1 s/1 s \\ \hline
    \end{tabular}
      \label{tab1}     
\begin{tablenotes}
\item Note that the separation angle between HXI and STIX is approximately 14 degrees. Data from GBM's detector-2 were used in the study.
\end{tablenotes}
\end{table}

WST observes the Sun in the Balmer continuum at 3600$\pm$20 \AA\ (hereafter the WST continuum) in two modes: a routine mode for full-disk imaging with a 2-minute cadence, and a burst mode for a flaring active region, featuring a high cadence over ten minutes with time resolutions of 1 s for the first five minutes and 2 s for the last five minutes. 
In this study, all WST images were rebinned to 1.5\arcsec\ per pixel.
The burst-mode imaging data of WST for the last five minutes were interpolated to a 1-second cadence, and other full-disk integrated data were also rebinned to 1 s to match with WST data. 
The vector magnetic field at 16:48 UT from the ``hmi.sharp\_cea\_720s'' series product of HMI was utilized in a nonlinear force-free field (NLFFF) extrapolation to analyze magnetic configurations \citep{Bobra2014}.  
During the flare, data gaps exist in both LYRA and EOVSA. Meanwhile, HXI and STIX were affected by Earth’s radiation belts and attenuator insertion, respectively. 

The Morlet wavelet transform was employed to determine QPP periods (see \citealt{TC1998}).
Within this methodology, the local 95\% confidence level for the wavelet power spectrum (WPS) was established through the following steps: 1) a red-noise background with a specified lag-1 autocorrelation was assumed for the observed data, 2) the theoretical wavelet power spectrum for this red-noise process was derived, and 3) the observed WPS was normalized and compared against the theoretical red-noise spectrum, with a chi-squared distribution then used to determine the 95\% confidence level.
We applied smoothing and detrending to all time profiles before conducting the wavelet analysis.
Note that smoothing was performed using the TS\_SMOOTH function in IDL (ts\_smooth.pro).
To avoid artificial periods caused by the detrending process, we followed the method proposed by \cite{Dominique2018} and tested different smoothing widths (SWs) within a range of 3--500 s.
The specific identification criteria are: 1) the period must be present in data from at least two different instruments, 
2) it must exceed the red noise level as determined by global wavelet significance, 
3) it should be less than or equal to half of the SW, and 
4) it must stably appear across varying SWs. 
The QPP periods met these criteria are listed in Table \ref{tab2}.

The Fourier transform was employed to determine the spatial distribution of QPP sources.
To determine magnetic structures of the active region, we performed a NLFFF extrapolation using the magneto-frictional method proposed by \citet{Guo2016} within the framework of the Message Passing Interface Adaptive Mesh Refinement Versatile Advection Code (MPI-AMRVAC; \citealt{Keppens2023}).

\section{Results}
\label{sec:result}

\subsection{Overview of the X2.8 white-light flare}

\begin{figure}[htb]
	\centering
	\includegraphics[width=0.57\textwidth]{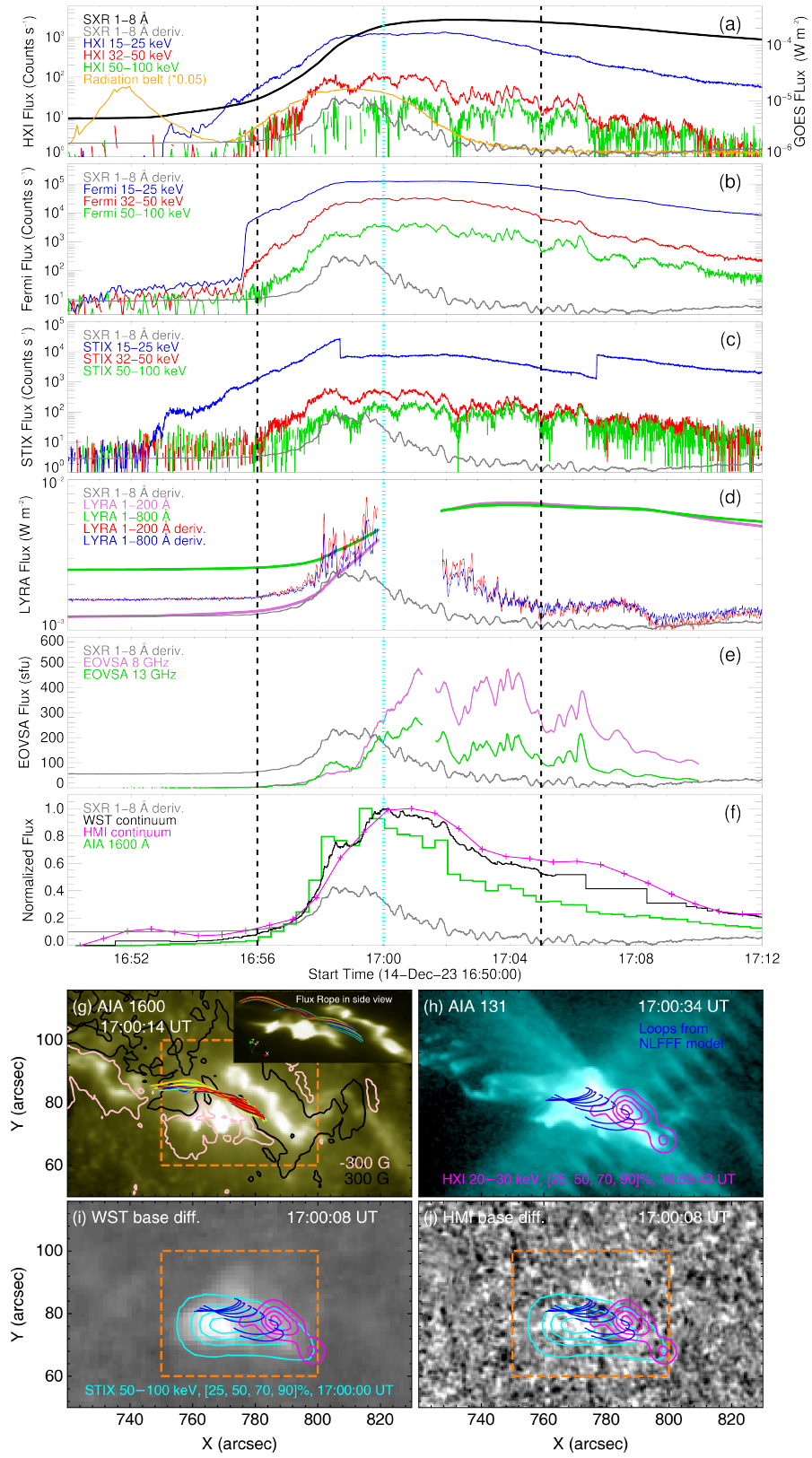}
	\caption{Overview of the X2.8 WLF. (a)--(e): Full-disk integrated time profiles from GOES/XRS, ASO-S/HXI, Fermi/GBM, SolO/STIX, PROBA2/LYRA, and EOVSA. All of the HXR curves are background-subtracted, and the influence of the radiation belt is illustrated by the orange line in panel (a). The pronounced reduction in STIX 15--25 keV is due to the impact of attenuator. (f): time profiles of the flaring region observed in HMI, WST, and AIA 1600 \AA. Their integration area is enclosed by the orange box in panels (g), (i), and (j). The vertical cyan and black lines denote the WST continuum peak time and high-cadence imaging interval, respectively.
	(g)--(j): Multiwavelength imaging of the flare around 17:00 UT, including AIA 1600 \AA, 131 \AA, and base difference maps of WST and HMI. Line-of-sight magnetic fields and a magnetic rope obtained from the NLFFF extrapolation are overplotted on panel (g). HXR sources from HXI and STIX, along with magnetic loops from the NLFFF extrapolation, are overlaid on panels (h)--(j).
	Note that the HXI source position may be spatially biased due to radiation belt contamination, but this bias is difficult to quantify.}
	\label{fig1}
\end{figure}

The X2.8 flare occurred in the NOAA 13514 active region (N04W53), beginning at 16:47 UT, peaking at 17:02 UT, and ending at 17:12 UT on 2023 December 14. 
Figure \ref{fig1} presents multiwavelength time profiles of the flare. 
We can see that high-energy HXR emissions (above 32 keV) persist for a long time during the flare, accompanied by pulsations with different time scales.
These HXR pulsations are also evident in microwave emissions and time derivatives of extreme ultraviolet (EUV)/SXR emissions, correlating with some minor peaks in the WST continuum curve.
The time profile of WST continuum peaked around 17:00 UT, approximately 35 s later than the AIA 1600 \AA\ profile and the time derivative of GOES 1--8 \AA.

Multiwavelength observations and the NLFFF extrapolation reveal that this is a typical two-ribbon flare, as shown in panels (g)--(j). 
Panel (g) illustrates two ribbons of the flare and a magnetic rope near the magnetic neutral line before flare onset. 
The flare arcade is well revealed by AIA 131 \AA\ imaging, closely resembling the magnetic loops extrapolated in the NLFFF approximation. 
The HXI 20--30 keV source is situated near the western flare ribbon and/or loop top, while the STIX 50--100 keV source is most spatially correlated with the eastern flare ribbon, where the white-light continuum enhancement is strongest.

\subsection{Time lags in the multiwavelength observations}
\label{sec:timelag}

\begin{figure}[htb]
	\centering
	\includegraphics[width=0.75\textwidth]{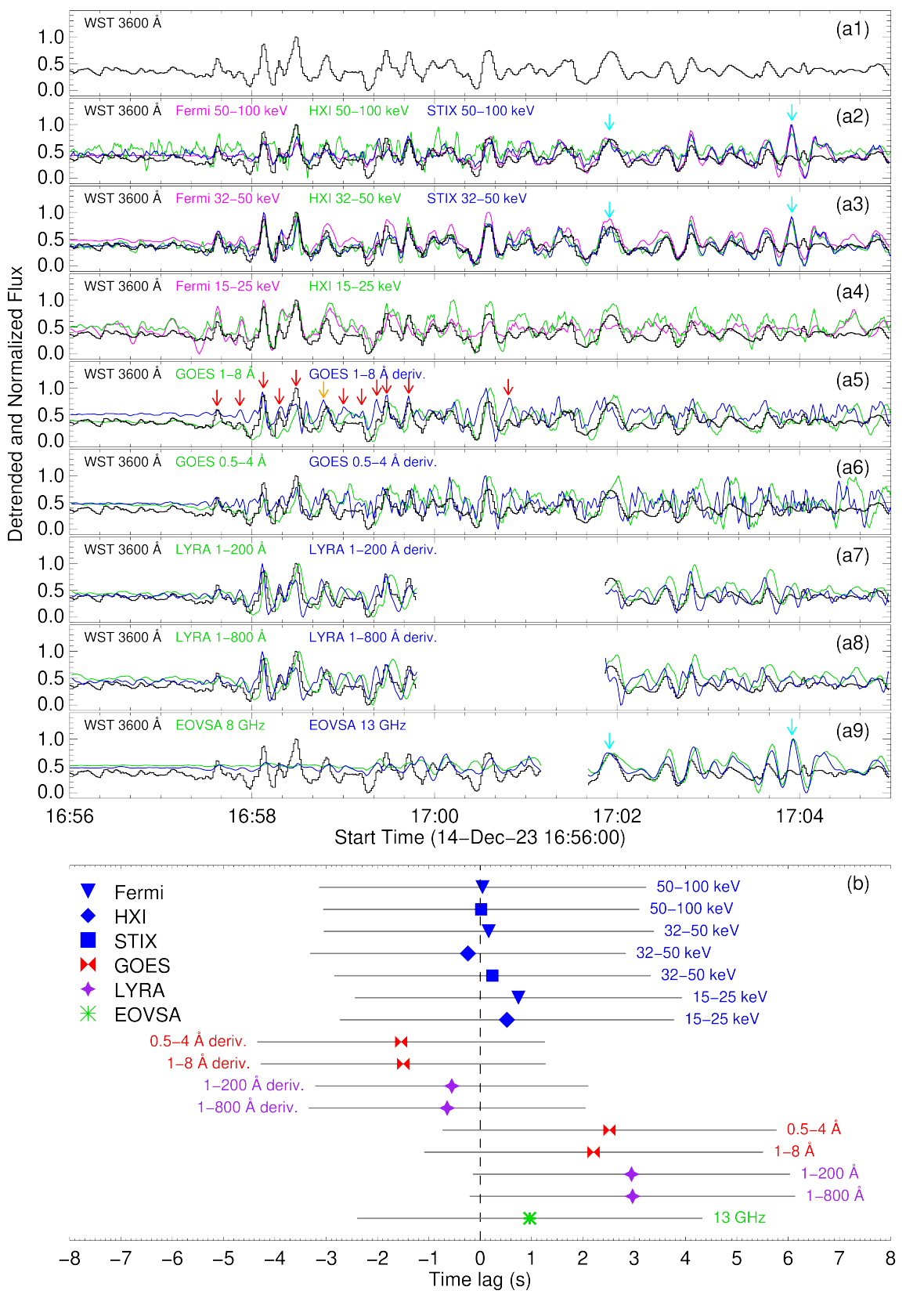}
	\caption{Analysis on time lags. (a1)--(a9): Detrended and normalized time profiles across multiple wavelengths in the WST high-cadence imaging interval. Panel (a1) displays the WST continuum, also overlaid on the panels (a2)--(a9). Cyan arrows in panels (a2), (a3), and (a9) indicate a quasi-simultaneous pulses in the HXR and microwave bands. Red arrows in panel (a5) mark pulses peaking closer to the WST continuum in the GOES 1--8 \AA\ derivative curve, while the single orange arrow marks a pulse peaking earlier in the same curve. (b): Averaged time lags for the time profiles in panels (a2)--(a9) relative to the WST continuum. Negative time lags indicate the waveband leads the WST continuum. Only bands with a correlation coefficient greater than 0.4 are shown. The error bar (gray horizontal line) is estimated as $\sqrt{\sigma ^{2}+{s_{max}}^{2}}$, incorporating both the standard deviation $\sigma$ derived from a single Gaussian fit to the cross-correlation function, and $s_{max}$, representing the maximum sampling time of the WST in the period, which is 2 s.}
	\label{fig2}
\end{figure}

The correlation of pulse peak timings across multiwavelength observations reflects energy release and deposition processes occurring during the flare.
To clearly present the pulses in high-cadence ($\leq$1 s) wavebands and compare their time lags, 2-second and 20-second SWs were successively applied to the time profiles to reduce the effect of noise and filter out their slowly varying components (similar to the approach of \cite{Dolla2012}), as shown in Figure \ref{fig2}.

From Figures \ref{fig2}(a1)--(a4), we can see that peak times in the WST continuum are highly consistent with those observed in the HXR emissions, which is also supported by the cross-correlation analysis results represented in Figure \ref{fig2} (b). 
Such consistency highlights that the enhancement of the Balmer continuum in the lower atmosphere is sensitive to the bombardment of non-thermal electrons, as previously reported \citep[e.g.][]{Dominique2018L,LiD2021,Joshi2021,LQ2024,LD2025AA}.
Additionally, minimal time delays in HXR emissions across different energy bands imply weak time-of-flight and solar atmosphere response effects \citep{Qiu2012,Dolla2012}.

Figures \ref{fig2}(a5)--(a8) illustrate the temporal relationship between SXR/EUV emissions and the WST continuum. Utilizing GOES 1--8 \AA\ observations as an exemplar and examining HXR data (Figures \ref{fig2}(a2)--(a4)), we can observe that a predominant proportion of SXR time derivative pulse peaks exhibit temporal coincidence with both WST continuum and HXR pulse peaks (red arrows, Figure \ref{fig2}(a5)). This finding is further supported by cross-correlation analysis (Figure \ref{fig2}(b)) and aligns with the Neupert effect \citep{Neupert1968}, which posits that non-thermal particles (specifically electrons) deposit energy in the chromosphere, leading to chromospheric evaporation \citep[e.g.,][]{LiD2023}. However, we also note that some of SXR time derivative pulse peaks precede the WST continuum and HXR pulses (e.g. the orange arrow in Figure \ref{fig2}(a5)).
This earlier occurrence of SXR time derivative pulses relative to HXR pulses, reported sporadically in previous studies \citep[e.g.,][]{Simoes2015,Hayes2016}, may relate to heating processes during the flare beyond non-thermal electron bombardment.
For this specific flare event, the GOES temperature profile peaks proximate to the time indicated by the orange arrow, implying thermal conduction as the dominant chromospheric heating mechanism at this time. It is important to note that other heating mechanisms cannot be excluded.

Additionally, as shown in Figure \ref{fig2}(a9), the microwave pulses observed by EOVSA become prominent after the white-light peak time ($\sim$17:00 UT), also depicted in Figure \ref{fig1}(e).
Many of these microwave pulses slightly lags behind the pulses observed in WST continuum and HXR bands, which can be interpreted by the trap-plus-precipitation model \citep{Minoshima2008}. 
We can also notice that certain microwave pulses exhibit strong alignment with HXR emissions, as indicated by the cyan arrows in Figures \ref{fig2}(a2), (a3), and (a9), reflecting the common origin of the non-thermal electrons responsible for producing bremsstrahlung and gyrosynchrotron emissions.

We also explored the relationships of time lags before and after the white-light peak time, finding no significant difference between the separated intervals and the entire interval of interest.

\subsection{Quasi-harmonic QPPs suggested by multiwavelength observations}

\begin{figure}[htb]
	\centering
	\includegraphics[width=0.9\textwidth]{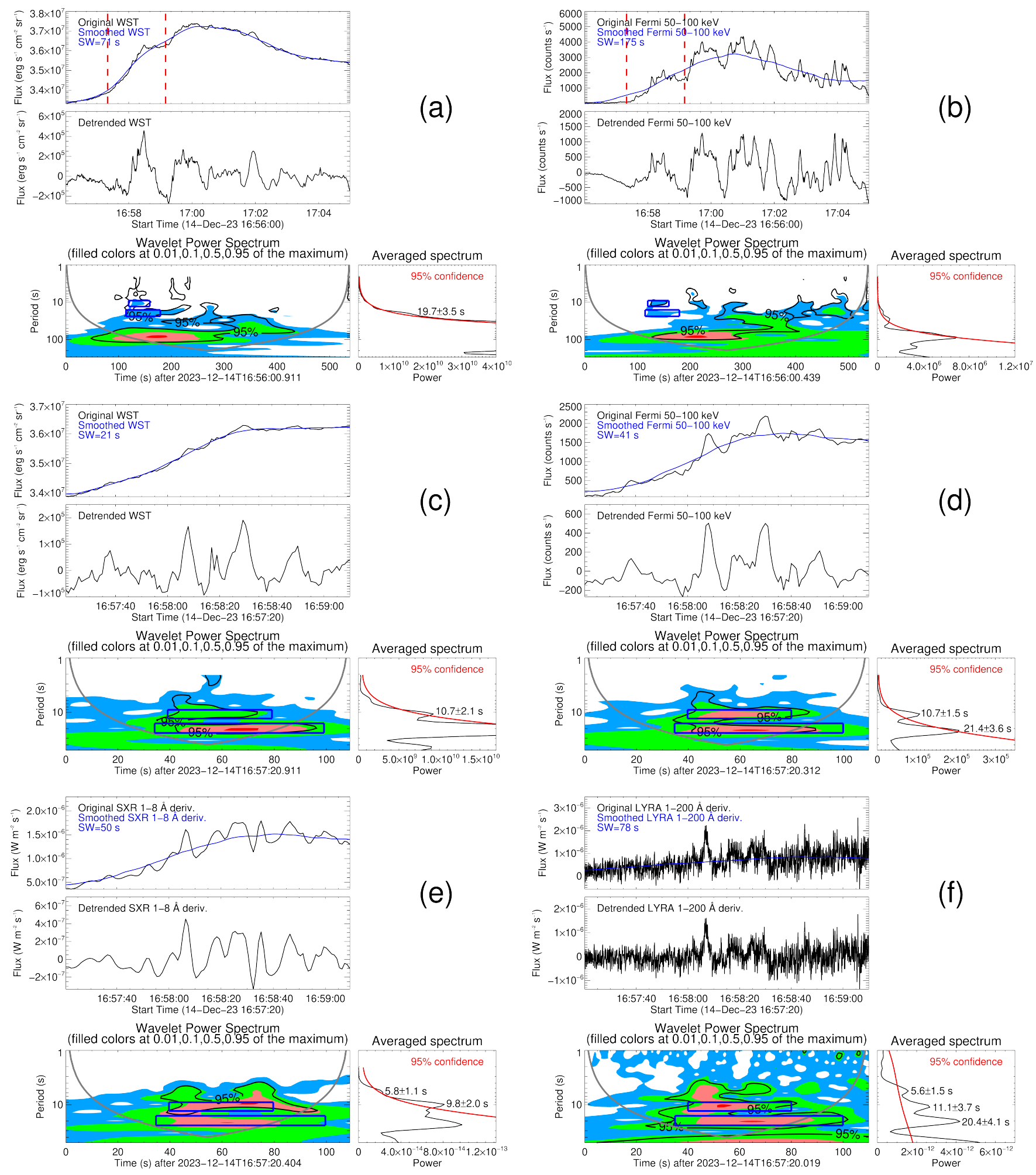}
	\caption{Wavelet analysis results. (a)--(f): Six sets of subplots include the wavelet analysis of WST continuum, Fermi 50--100 keV, and time derivatives of GOES 1--8 \AA\ and LYRA 1--200 \AA. Each set displays: original and smoothed curves (top left), detrended curves (middle left), wavelet power spectrum (WPS; bottom left), and time-averaged global WPS (bottom right). The black thick contour in the WPS diagrams encloses regions of exceeding 95\% confidence against a red-noise background, and the gray curve indicates the cone of influence. Smaller and larger blue rectangles represent the dominant domains in the time interval and period range for the 11$\pm$2 s and 20$\pm$4 s oscillations, respectively. The global 95\% confidence level is marked by a red line in each time-averaged WPS, along with significant QPP periods and their uncertainties. The uncertainty is estimated using the peak half-widths of the global WPS. Panels (a)--(b) correspond to the results for the period of interest, while panels (c)--(f) correspond to the results for the subinterval, delimited by the red dashed lines in panels (a) and (b).}
	\label{fig3}
\end{figure}

This flare exhibited signatures of quasi-harmonic QPPs during its impulsive phase, particularly in the rising phase of the white-light continuum, approximately from 16:58 to 16:59 UT.
Some representative wavelet analysis results are shown in Figure \ref{fig3}. 
As illustrated in Figure \ref{fig3}(a), a subtle 20-second QPP was suggested around 16:58--16:59 UT, enclosed by the larger blue rectangle in its WPS diagram.
This 20-second period also displays a corresponding enhancement in the WPS of Fermi 50--100 keV energy band, even with a large SW of 175 s, although this enhancement does not dominate in its WPS diagram (see the larger blue rectangle in Figure \ref{fig3}(b)). 

Figures \ref{fig3}(c)--(f) further illustrate the possible QPP periods of 20 s and shorter in the rapid rising phase of the WST continuum, which is delimited by the red vertical dashed lines in Figures \ref{fig1}(a) and (b).
Specifically, a QPP period of approximately 11 s is present in the WST continuum, Fermi 50--100 keV, and the time derivatives of GOES 1--8 \AA\ and LYRA 1--200 \AA\ (see Figures \ref{fig3}(c)--(f)). 
The 20-second QPP mentioned earlier is more prominently appeared during the rising phase in both the Fermi 50--100 keV and the time derivative of LYRA 1--200 \AA, as shown in Figures \ref{fig3}(d) and (f).
Notably, the time derivative of the 1--200 \AA\ band reveals an even shorter period of approximately 6 s (see Figure \ref{fig3}(f)). This period, however, is very weak in the time derivative of 1--8 \AA\ band and absent in WST continuum and HXR emissions.

Therefore, considering multiwavelength observations collectively, we infer the presence of quasi-harmonic QPPs with a fundamental period of $\sim$20 s and a second harmonic of $\sim$11 s. See Table \ref{tab2} for details.

\begin{table}[htbp]
  \centering
  \caption{Quasi-harmonic QPPs implied by flare-integrated multiwavelength observations}
    \begin{tabular}{l|ccc}
		\hline
		\hline
    \multirow{4}{*}{\diagbox[width=5.5cm,height=4\line,dir=SE,innerleftsep=-3.6em]{Wave band}{QPP period\\ \& Time interval}} 
   &  $\sim$6    & $\sim$11    & $\sim$20     \\
       & (s)   & (s)    & (s)    \\
\cline{2-4}          
  & \multicolumn{3}{c}{16:57:20--16:59:10}  \\
    &  \multicolumn{3}{c}{(UT)}  \\ 
    \hline
    3600$\pm$20 \AA\ &       &   $\circ$ &  $\circ$  \\
    \hline
    HXR 50--100 keV &       &$\checkmark$ & $\circ$   \\
    \hline
    HXR 32--50 keV &       &$\checkmark$ & $\checkmark$   \\
    \hline
    HXR 15--25 keV &       &    $\checkmark$   &       \\
    \hline
    1--8 \AA\ time derivative &     $\circ$   &   $\checkmark$       \\
    \hline
    1--200 \AA\  time derivative &        $\checkmark$   &    $\checkmark$   &    $\checkmark$   \\
    \hline
    1--800 \AA\ time derivative  &         $\checkmark$   &    $\checkmark$   &    $\checkmark$     \\
    \hline
    13 GHz &       &     $\checkmark$    &      \\
    \hline
    \hline
    \end{tabular}%
  \label{tab2}
\begin{tablenotes}
 \item Note that QPP periods satisfying the criteria described in Section \ref{instr_data} are marked by symbols: $\checkmark$ for prominent, and $\circ$ for weak.
 \end{tablenotes}
\end{table}%

\subsection{Spatial distribution of the quasi-harmonic QPP sources}
\label{SpQPP}
 
\begin{figure}[htb]
	\centering
	\includegraphics[width=1.0\textwidth]{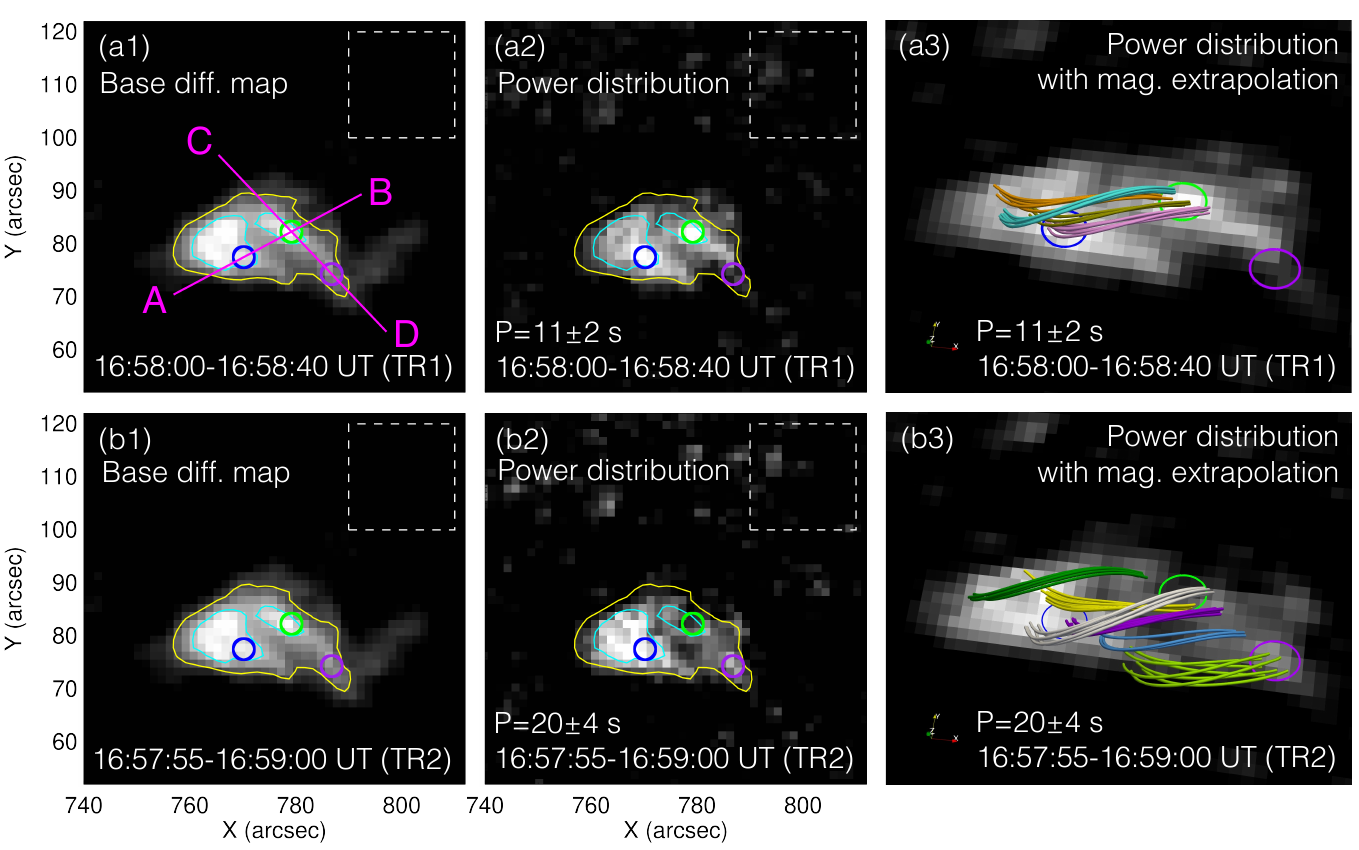}
	\caption{Spatial distribution of the QPP sources around the periods of 11 s and 20 s. Panel (a1) presents the temporal-averaged base difference map over the dominant time interval (i.e. TR1) of the 11-second QPP, while panel (a2) displays the temporal-and-frequency-averaged Fourier power map for the same interval. Its dominant domain in time and frequency is enclosed by the smaller blue rectangles in the WPS diagrams in Figure \ref{fig3}. In panels (a1) and (a2), pixels are displayed only where intensity values surpass three standard deviations (3$\sigma$) of the background fluctuations. The background region, situated to the northwest of the flare, is enclosed by the white dashed box in panels (a1), (a2), (b1), and (b2). Cyan and yellow contours outline the flare ribbons. Blue, green, and purple circles enclose three regions of interest. The two slices AB and CD, indicated in panel (a1), are used to study the separation and elongation motions of  flaring ribbons as shown in Figures \ref{fig5}(a) and (b), respectively. (a3): Zoomed-in 3D view of the power map from panel (a2), illustrating magnetic loops associated with the green circle. Panels (b1)--(b3) have the same annotations as (a1)--(a3) but correspond to the 20-second QPP.}
	\label{fig4}
\end{figure}

Utilizing the high-cadence imaging data acquired by WST, Fourier Transform analysis was performed for every pixel to determine the spatial distribution of the quasi-harmonic QPP sources, following the methodology in previous studies \citep[e.g.][]{Yuan2019,LD2024}.
The dominant temporal and spectral domains of the quasi-harmonic QPPs can be broadly identified from the wavelet analysis results shown in Figures \ref{fig3}(c)--(f). Specifically, period ranges of 11$\pm$2 s ($\sim$0.111--0.077 Hz) and 20$\pm$4 s ($\sim$0.063--0.042 Hz) dominate the time intervals 16:58:00--16:58:40 UT (TR1) and 16:57:55--16:59:00 UT (TR2), respectively. 
The blue rectangles in the WPS diagrams of Figure \ref{fig3} visually represent these two dominant domains.
The temporal-averaged base difference maps and the temporal-frequency-averaged spatial power maps of these two QPPs are presented in Figure \ref{fig4}.

For the temporal-averaged emission enhancement (Figures \ref{fig4}(a1) and (b1)), the increase in the WST continuum of flare ribbons showed minimal spatial variation over TR1 (Figure \ref{fig4}(a1)) and TR2 (Figure \ref{fig4}(b1)), with the east flare ribbon being the dominant contributor.
As anticipated, the east ribbon simultaneously contributes the majority of the oscillation power for both 11-second and 20-second QPPs, as shown in Figures \ref{fig4}(a2) and (b2).
Furthermore, we notice that certain regions where the enhancement is not very strong also contribute notable power to both the QPP periods. For example, the region enclosed by blue circles in Figure \ref{fig4}.
More interestingly, a northern area of the west ribbon marked by green circles, which shows an obvious continuum enhancement, contributes significantly to the 11-second QPP but exhibits minimal contribution to the 20-second QPP. 
Conversely, a southern part of the west ribbon with weaker continuum enhancement, enclosed by purple circles, contributes more amplitude power to the 20-second QPP while offering little to the 11-second QPP. 

The NLFFF extrapolation reveals the magnetic connectivity between the two ribbons, particularly within three interesting circular regions, as illustrated in Figures \ref{fig4}(a3) and (b3).
We can see that the region exhibiting the strongest 11-second QPP power on the west ribbon is connected to multiple locations on the east flare ribbon, primarily within and north of the blue circular region (Figure \ref{fig4}(a3)).
Conversely, the purple circle and its northern region are mainly connected to the southern, interior, and surrounding areas of the blue circle on the east flare ribbon (Figure \ref{fig4}(b3)).

\subsection{Relation between the quasi-harmonic QPPs and flare ribbon evolution}

\begin{figure}[htb]
	\centering
	\includegraphics[width=1.0\textwidth]{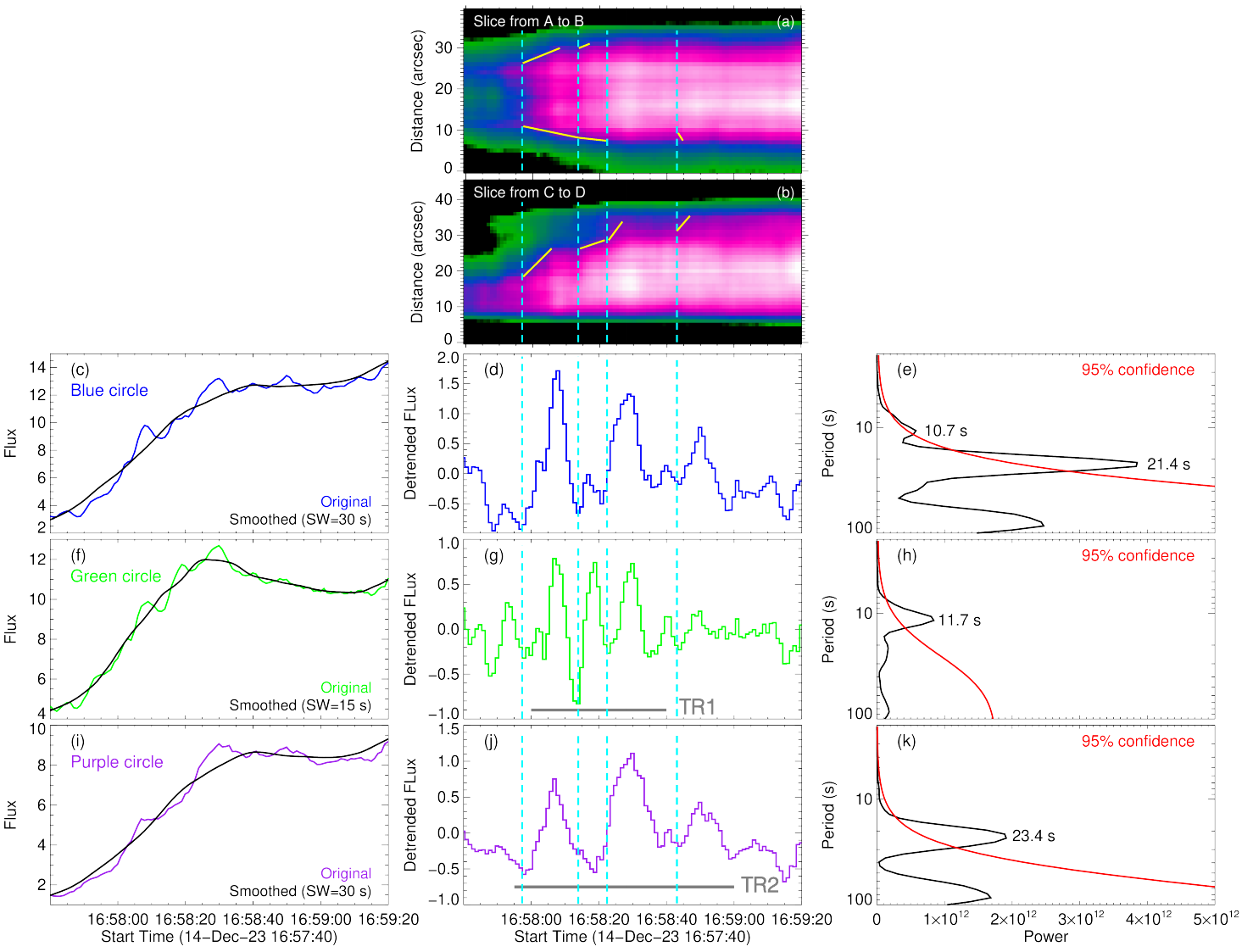}
	\caption{Evolution of flaring ribbons and the quasi-harmonic QPPs. Panels (a) and (b) show the time-distance diagrams of WST continuum along AB and CD (marked in Figure \ref{fig4}(a1)), respectively. Yellow lines in panels (a) and (b) indicate the separation and elongation motions of the flare ribbons, respectively. The bottom three rows display wavelet analysis results for the three circled regions in Figure \ref{fig4}, showing, from left to right: original and smoothed light curves, detrended light curves, and time-averaged global WPS. It is noted that the flux unit is $\text{erg s}^{-1}\text{ cm}^{-2}\text{ sr}^{-1}$. QPP periods for each of the three regions are indicated in the rightmost column panels. Vertical cyan dashed lines in the middle column panels indicate the onset of rapid flare ribbon evolution. Time intervals TR1 and TR2 are indicated by gray horizontal lines in panels (g) and (f).}
	\label{fig5}
\end{figure}

To further investigate the relationship between the quasi-harmonic QPPs and flare ribbon evolution, we constructed two time-distance (TD) maps to analyze the separation (Figure \ref{fig5}(a)) and elongation (Figure \ref{fig5}(b)) motions of the flare ribbon. The positions of the corresponding cut slices (AB and CD) are indicated in Figure \ref{fig4}(a1). Additionally, wavelet analysis was performed on the WST light curves within the three circled regions in Figure \ref{fig4}, with results presented in Figures \ref{fig5}(c)--(k).

Comparison of the TD maps and wavelet analysis results in Figure \ref{fig5} reveals that the occurrence of quasi-harmonic QPPs corresponds to rapid elongation and separation motions of the flare ribbons, as indicated by vertical cyan dashed lines in the middle column panels. 
Furthermore, the wavelet analysis results from the three circled regions are presented in the rightmost column of Figure \ref{fig5}.
It can be seen that the quasi-harmonic QPPs ($\sim$21 s and $\sim$ 11 s) are found in the blue-circled region (Figure \ref{fig5}(e)), while for the 
green- and purple-circled regions, they are predominantly dominated by QPPs with periods of $\sim$12 s and $\sim$23 s, respectively. 
The modulation depths, defined as the ratio of oscillatory amplitude to the long-term trend, for the $\sim$12 s and $\sim$23 s QPPs are approximately 6\%--9\% and 5\%--17\%, respectively. It is important to note that modulation depths vary across different regions.
These results align with the QPP power distribution maps shown in Figure \ref{fig4} and further reinforce the existence of quasi-harmonic QPPs in this WLF.

\section{conclusions and discussions}
\label{sec:summary}

In this study, we present the spatiotemporal properties of QPPs in an X2.8 two-ribbon solar WLF.
For the first time, we identify quasi-harmonic QPPs in the 3600 \AA\ continuum of a solar WLF and spatially pinpoint their QPP sources.
The main findings are as follows:

1. The enhancement of the WLF in both the Balmer and Paschen continua shows strong spatiotemporal correlation with HXR emissions, with the continuum emission peaking approximately 35 s later than the time derivative of the GOES 1--8 \AA\ emission.

2. During the flare, pulses in the WST continuum displayed a near-zero time lag with most pulses in HXR emissions and the time derivatives of SXR and EUV emissions.
However, the pulses in the EUV (1--800 \AA) and SXR (0.5--4, 1--8, 1--200 \AA) emissions lagged the WST continuum pulses by approximately 2--3 s.

3. Quasi-harmonic QPP periods of $\sim$11 and $\sim$20 s, initially suggested by multiple wavelengths in the rising phase of the white-light continuum, were further confirmed by WST imaging observations.

4. The 11-second and 20-second QPPs mainly originated from the east flare ribbon, which exhibited the strongest continuum enhancement. The west ribbon contributed significantly to the 11-second QPP but had a weaker contribution to the 20-second QPP.

5. The occurrence of quasi-harmonic QPPs is temporally coincident with the rapid elongation and separation motions of flare ribbons.

Harmonic QPPs have been reported in flares of the Sun and other Sun-like stars, often associated with the modulation of flare loops by MHD waves \citep[e.g.,][]{InglisNakariakov2009,Mancuso2020}.
In the X2.8 flare under study, QPPs with periods of approximately 11 s and 20 s exhibit characteristics of harmonically oscillating modes modulated by MHD waves, with 20 s as the fundamental mode and 11 s as the second harmonic, respectively.

Existing observations allow us to exclude certain QPP mechanisms.
For instance, this flare did not exhibit significant QPPs in thermal emissions. 
Thus, we can largely exclude mechanisms involving strong compressibility or significant modulation of thermal processes, such as sausage modes and the thermal instability in the current layer, etc \citep[see][]{Zimovets2021}. 
Besides, from the multiwavelength imaging, we can further discount several other mechanisms, such as reconnection triggered by outer MHD waves and dispersive wave trains \citep[e.g.,][]{Nakariakov2006,Pascoe2017b,Zhou2024}.

Utilizing observations from multiple instruments, e.g. AIA and GOES, the spectral fitting of STIX, and NLFFF extrapolation results, 
we roughly estimated the physical parameters of the flare loops. 
These parameters include the loop length $L$, temperature $T$, magnetic field strength around the loop top $B$, and plasma density inside the loop $n_{i}$, estimated to be approximately 14--29 Mm, 7.1--32.7 MK, 460--1250 G, and (0.2--9.0)$\times 10^{11} \mathrm{~cm}^{-3}$, respectively.
Detailed estimations of $T$ and $n_{i}$ are given in Appendix \ref{appendixA}.

Subsequently, we can assess the potential modulations by slow-mode and kink-mode waves.
For slow-mode waves, the period is given by the formula $P_{slow}=\frac{2 L}{j c_{T}}$, where $j$ is the oscillation mode number, and $c_{T}$ is the slow magnetoacoustic speed, which is close to the sound speed in typical coronal loops \citep{Roberts1984,Mariska2006}.  
Given the magnetoacoustic speed $c_{s} \approx 0.152 T^{1/2} \mathrm{~km} \mathrm{~s}^{-1}$, the fundamental slow mode ($j=1$) predicts a QPP period of 32.2--143.2 s, exceeding the observed fundamental 20-second period. 
For kink-mode waves, the period is determined by the formula $P_{kink}=\frac{2L}{jc_{Ai}(\frac{2}{1+n_{e}/n_{i}})^{1/2} }$, where $c_{Ai}$ is the \alfven\ speed inside the loop and $n_e$ is the plasma density outside the loop. The \alfven\ speed is approximated as $c_{Ai}\approx \frac{2.18B}{{n_{i}}^{1/2}}\times 10^{11} \mathrm{~cm} \mathrm{~s}^{-1}$.
Assuming the density ratio $n_{e}/n_{i}\approx ~$0.1--0.5 \citep{Aschwanden2003}, the range of QPP periods modulated by the fundamental kink-mode wave is 1.3--40.7 s, which includes the detected fundamental 20-second period. 
Thus, kink-mode waves are more likely than slow-mode waves to modulate the non-thermal emissions of the flare.

Additionally, advanced numerical simulations have provided potential scenarios for QPPs in flares. Specifically, some recent numerical models suggest that charged particles can be effectively accelerated within the magnetic bottle at loop tops, leading to periodic oscillations in density and velocity, which can account for many observational features seen in two-ribbon flares \citep[e.g.,][]{Takahashi2017,Zhao2019}. 
In our event, the generation of quasi-harmonic QPPs is accompanied by the rapid elongation and separation motions of the flare ribbons. Such motions are widely recognized as being directly related to the flare magnetic reconnection process \citep{Qiu2017,Yang2025}, as predicted by the two-ribbon flare model \citep{PriestForbes2000}. Our observations suggest a strong correlation between the observed quasi-harmonic QPPs and the magnetic reconnection process inherent to flares. 

Recently, \citet{Kou2022} reported the simultaneous detection of 10--20 s QPPs in both EOVSA microwave (8.4 GHz) and the time derivative of GOES SXR 1--8 \AA\ observations. These QPPs were evident in both the flare region and the overlying current sheet. Combined with 2.5D MHD simulations, their study demonstrated that these non-thermal QPPs can be attributed to quasi-periodic magnetic reconnection modulated by magnetic island formation within the flare current sheet. Similarly, in our event, simultaneous non-thermal QPP signals with a similar period of $\sim$11 s is also observed in both EOVSA microwave (13 GHz) and the GOES SXR 1--8 \AA\ time derivative. Unfortunately, EOVSA imaging of this X2.8 flare was not feasible due to some reasons (Sijie Yu, private communication). Nevertheless, this similarity may still imply that analogous underlying physical processes are involved in the flare under study.

In conclusion, our observations generally support a quasi-periodic modulation of the efficiency of non-thermal electron production. This modulation could arise from oscillatory regimes inherent to magnetic reconnection, or alternatively, via oscillatory modulation of reconnection by MHD waves in the flare loops, among which the kink mode is the most consistent with our observations. Based on the available observational evidence, we cannot definitively distinguish between these possibilities for the X2.8 flare.
Furthermore, due to observational limitations, we can not exclude the possibility of self-oscillatory processes, such as reconnection triggered by MHD auto-wave processes.

QPPs on the timescale of 10--20 seconds have been reported in stellar WLFs, albeit without spatial resolution \citep[e.g.,][]{Mathioudakis2006,Tsap2011}.
In our study, the spatiotemporal characteristics of harmonic QPPs revealed by WST imaging in the Balmer continuum provide valuable insights for modeling solar and stellar WLFs \citep[e.g.,][]{Nizamov2019,Song2023}.
Our analysis not only improves the understanding of the physical mechanisms of QPPs under spatially resolved scenes but also constrains the heating functions in WLF modeling, thereby contributing to the understanding of energy transfer processes in WLFs.
Especially, the spatial distribution of the QPP sources identified through WST imaging can serve as critical references for future three-dimensional modeling of WLFs.

\acknowledgments
We thank the anonymous reviewer and the statistics editor for their constructive comments that improved the quality of our paper.
The ASO-S mission is supported by the Strategic Priority Research Program on Space Science, Chinese Academy of Sciences.
SDO is a mission of NASA's Living With a Star Program. 
The wavelet software, developed by C. Torrence and G. Compo, can be accessed at \url{http://atoc.colorado.edu/research/wavelets/}.
We thank Dr. Laurent Dolla, Dr. Sijie Yu,,Wenhui Yu, and Zhichen Jing for helpful discussions. 
Especially, we thank Sijie Yu for helping to check the availability of the EOVSA imaging data of the flare under study.
We also thank the teams of SolO, PROBA2, GOES, Fermi, and EOVSA for their open data use policy.
This work is supported by the Strategic Priority Research Program of the Chinese Academy of Sciences under grant XDB0560000, NSFC under grants 12273115, 12233012, and 12333010, and National Key R\&D Program of China under grant 2022YFF0503004, Natural Science Foundation of Jiangsu Province BK20241707.
De-Chao Song and Qiao Li are supported by the EUl Guest Investigator Program from the Belgian Federal Science Policy Office (BELSPO).
M. D. acknowledge support BELSPO in the framework of the ESA-PRODEX program, PEA 4000145189.
I. Zimovets and B. Nizamov are supported by a grant from the Russian Science Foundation (project no. 20-72-10158).
A. F. Battaglia is supported by the Swiss National Science Foundation Grant 200020\_213147.
De-Chao Song is also supported by the Jiangsu Funding Program for Excellent Postdoctoral Talent.

\appendix
\section{Estimation of temperature and plasma density inside the flare loop}
\label{appendixA}
STIX imaging and spectroscopy were utilized to estimate the density of thermal flare loops during the primary interval of the quasi-harmonic QPPs, specifically TR2 (16:57:55--16:59:00 UT). This approach was adopted by three main factors: (1) some AIA images were saturated during TR2, preventing parameter estimation (e.g., temperature and density) via the differential emission measure method; (2) HXI data were influenced by the Earth's radiation belts during TR2, affecting the spectral fitting results; and (3) STIX can acquire lower-energy X-ray data than HXI, which is important for thermal plasma parameter estimation.

\begin{figure}[htb]
	\centering
	\includegraphics[width=1.0\textwidth]{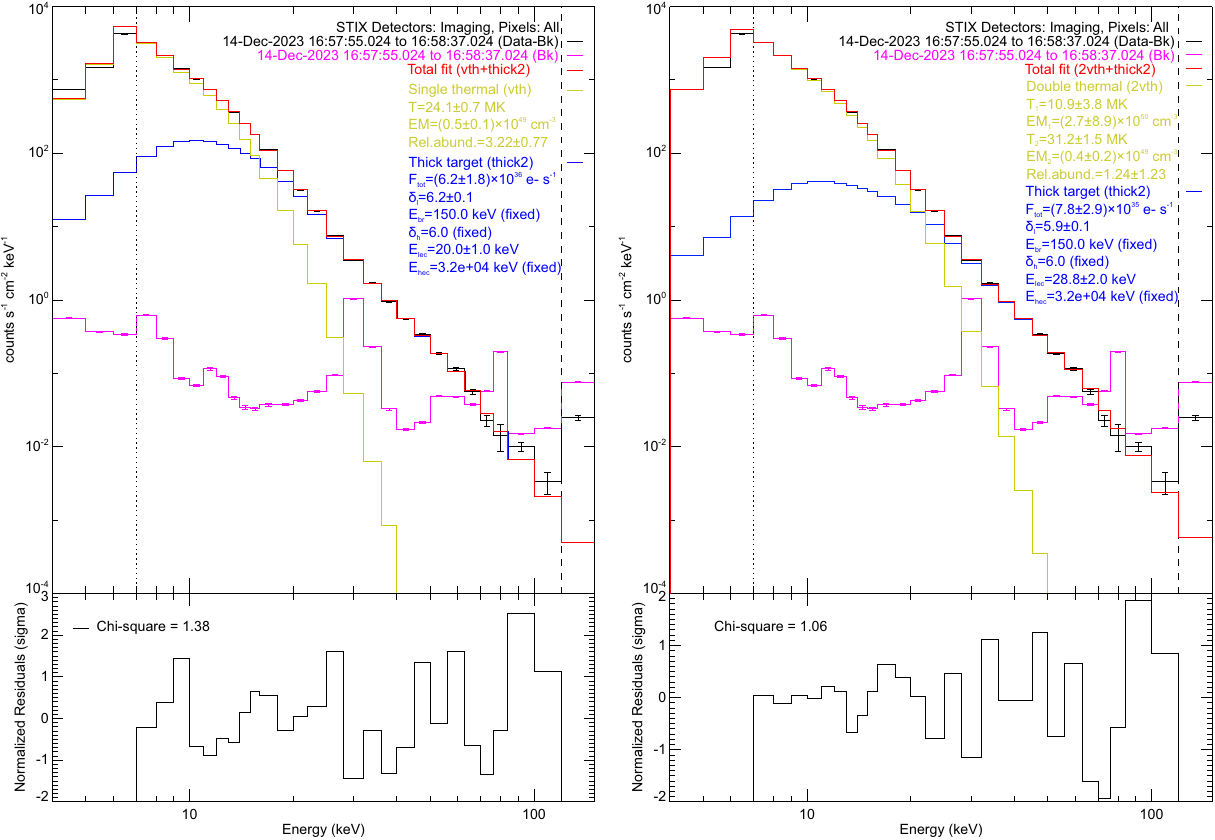}
	\caption{STIX spectral fittings. The left panel presents the STIX spectral observations recorded between 16:57:55 UT and 16:58:37 UT, along with the fitting results based on a single thermal plus thick target model. Fitting residuals and corresponding chi-squared values are presented in the panel below. The fitting range spans from the vertical dotted line (7 keV) to the vertical dashed line (120 keV). The fitting parameters annotated in the upper right corner are: $T$ (plasma temperature), $EM$ (emission measure), Rel. abund. (relative abundance of Iron/Nickel, Calcium, Sulfur, and Silicon relative to the coronal abundance in Chianti), $F_{tot}$ (total integrated electron flux), $\delta_{l}$ (index of the electron flux distribution function below the break), $E_{br}$ (break energy, fixed), $\delta_{h}$ (index of the electron flux distribution function above the break, fixed), $E_{lec}$ (low energy cutoff), and $E_{hec}$ (high energy cutoff, fixed). The right panel shows the fitting results over the same time interval and spectral range using a double thermal plus thick target model.}
	\label{STIX_fitting}
\end{figure}

Both the single thermal plus thick target model (vth+thick2) and the double thermal plus thick target model (2vth+thick2) were used for spectral fitting, as shown in Figure \ref{STIX_fitting}. We can notice that comparable fitting qualities were obtained for both models in terms of chi-squared values (1.38 for vth+thick2 and 1.06 for 2vth+thick2). However, the 2vth model yielded a more plausible coronal abundance (1.24$\pm$1.23), closer to 1, compared to the vth model (3.22$\pm$0.77). Thus, fitting results of the 2vth+thick2 model were employed to estimate the average density of flare loops, resulting in $EM\approx$(0.002--1.16)$\times 10^{51} \mathrm{~cm}^{-3}$ and $T\approx$7.1--32.7 MK. Note that GOES measurements for EM and T during TR2 fall within the ranges estimated by STIX.

During TR2, STIX imaging at lower energy bands (e.g., 4--5 keV and 9--14 keV) approximated a circular shape in the sky plane with a radius $R\approx$7--10 Mm. Under the assumption that the flare loop volume is approximated by a sphere with this circular area as a cross-section, the volume $V$=$\frac{4}{3}\pi R^{3}\approx$(1.4--4.2)$\times 10^{27} \mathrm{~cm}^{3}$ was derived. Finally, the plasma density of the flare loop ($n_{i}$) was estimated to be approximately (0.2--9.0)$\times 10^{11} \mathrm{~cm}^{-3}$. 
It should be noted that STIX fittings using other energy ranges (e.g., 5 or 6 keV to 120 keV) were also tested, yielding comparable parameter ranges.

\bibliographystyle{apj}

\end{document}